\documentclass[aps,prl,preprint,superscriptaddress]{revtex4}

\usepackage{graphicx}
\usepackage{dcolumn}
\usepackage{bm}
\usepackage{amsmath}
\usepackage{amssymb}
\usepackage{latexsym}
\usepackage{epsfig}
\usepackage{amsbsy}
\usepackage{array}
\usepackage{amssymb}
\usepackage{setspace}
\usepackage{bm}

\def\sint{\ifmmode{- \!\!\!\!\!\! \int}
    \else{\hbox{$- \!\!\!\! \int \ $}}\fi}

\begin{document}


\title{All-oxide spin Seebeck effects}

\author{Z. Qiu\footnote{Author to whom correspondence should be
addressed; electronic mail: qiuzy@imr.tohoku.ac.jp}}
\affiliation{WPI Advanced Institute for Materials Research, Tohoku University, Sendai 980-8577, Japan}
\affiliation{Spin Quantum Rectification Project, ERATO, Japan Science and Technology Agency, Sendai 980-8577, Japan}

\author{D. Hou}
\affiliation{WPI Advanced Institute for Materials Research, Tohoku University, Sendai 980-8577, Japan}
\affiliation{Spin Quantum Rectification Project, ERATO, Japan Science and Technology Agency, Sendai 980-8577, Japan}

\author{K. Uchida}
\affiliation{Spin Quantum Rectification Project, ERATO, Japan Science and Technology Agency, Sendai 980-8577, Japan}
\affiliation{Institute for Materials Research, Tohoku University, Sendai 980-8577, Japan}
\affiliation{PRESTO, Japan Science and Technology Agency, Saitama 332-0012, Japan}

\author{E. Saitoh}
\affiliation{WPI Advanced Institute for Materials Research, Tohoku University, Sendai 980-8577, Japan}
\affiliation{Spin Quantum Rectification Project, ERATO, Japan Science and Technology Agency, Sendai 980-8577, Japan}
\affiliation{Institute for Materials Research, Tohoku University, Sendai 980-8577, Japan}
\affiliation{Advanced Science Research Center, Japan Atomic Energy Agency, Tokai 319-1195, Japan}


\begin{abstract}
We report the observation of longitudinal spin Seebeck effects (LSSE) in an all-oxide bilayer system comprising an IrO$_2$ film and an Y$_3$Fe$_5$O$_{12}$ film. Spin currents generated by a temperature gradient across the IrO$_2$/Y$_3$Fe$_5$O$_{12}$ interface were detected as electric voltage via the inverse spin Hall effect in the conductive IrO$_2$ layer. This electric voltage is proportional to the magnitude of the temperature gradient and its magnetic field dependence is well consistent with the characteristic of the LSSE. This demonstration may lead to the realization of low-cost, stable, and transparent spin-current-driven thermoelectric devices. 
\end{abstract}



\maketitle

The spin Seebeck effect (SSE) generates spin voltage in a magnetic material as a result of a temperature gradient \cite{Weiler2012,Uchida2008,Jaworski2010,Qu2013, Agrawal2014,Rezende2014, R.Ramos2013,Meier2013,Uchida2010,Uchida2010a,Uchida2012,Kikkawa2013,Kikkawa2013a,Kirihara2012, Roschewsky2014,Uchida2014a,PhysRevX.4.041023,Schreier2013a,Uchida2013a}. Since the thermally generated spin voltage induces a spin current across the interface between the magnetic material and an adjacent conductive material, it can be detected as electric voltage via the inverse spin Hall effect (ISHE) in the conductive layer. Therefore, a magnetic/conductive bilayer system is commonly used for the SSE study \cite{Uchida2010,Uchida2010a,Kirihara2012,Uchida2012,Kikkawa2013,Kikkawa2013a,R.Ramos2013,Meier2013, Roschewsky2014,Uchida2014a,PhysRevX.4.041023,Schreier2013a,Uchida2013a}. The SSE has attracted increasing attention because of the possible applications for thermoelectric conversion and spintronic devices \cite{Uchida2014a,Bauer2012}, and investigation of the SSE in various materials is important to further improvement of thermoelectric and thermo-spin conversion efficiency. However, all the experimental studies on the SSE to date have been performed using simple metals as conductive layers, while widen variety of materials have been investigated for the magnetic layer\cite{R.Ramos2013,Meier2013,Uchida2013a}.

As alternate conductive materials for the SSE devices, conductive oxides can be good candidates because of the low cost, good chemical stability, and easy preparation of oxides. In addition, conductive oxide films can often be transparent. Therefore, the conductive oxides enable the construction of transparent thermoelectric and thermo-spin devices, making it more suitable for applications requiring transparency, such as applications to smart windows.

In this work, we report the observation of the SSE in an all-oxide bilayer system comprising a conductive IrO$_2$ and a ferrimagnetic insulator Y$_3$Fe$_5$O$_{12}$ (YIG). Here, YIG is one of most widely used materials for spin-current studies since it has a small Gibert damping constant, long spin-wave-propagation length, and high electrical resistivity \cite{Kikkawa2013,Kikkawa2013a,Roschewsky2014,Uchida2014a,PhysRevX.4.041023,Uchida2010a,Uchida2010,Uchida2012,Kirihara2012, Geller1957,Cherepanov1993,Serga2010}. We select IrO$_2$ for detecting the SSE since relatively-high spin-Hall angle has been reported in this conductive oxide \cite{Qiu2013a,Qiu2012,Fujiwara2013}. Importantly, IrO$_2$ is an n-type transparent semiconductor of which its work function is very close to metals, such as Ag \cite{Chalamala1999}, making it easier to form an Ohmic contact between IrO$_2$ and metallic electrodes \cite{Fujiwara2013}. This situation is indispensable for detecting the ISHE in IrO$_2$ if the output voltage is in the order of submicrovolts or less.

\begin{figure}
\centering
\begin{minipage}[b]{0.5\textwidth}
\centering
\includegraphics[width=3in]{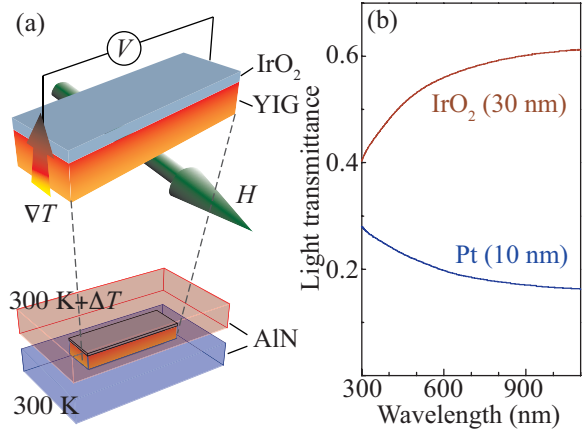}
\end{minipage}
\caption{(a) A schematic illustration of the IrO$_2$/YIG sample and experimental configuration for measuring the LSSE. $\nabla T$ and $H$ denote the temperature gradient and the external magnetic field, respectively. The sample was sandwiched between two AlN heat baths, of which the temperatures were stabilized to $300~\textrm{K} + \Delta T$ and 300 K, respectively. (b) Comparison of light transmittance spectra of a 30-nm-thick IrO$_2$ film and a 10-nm-thick Pt film. The films were formed on glass substrates, and the contribution from the light transmittance of the substrates is subtracted. 
\label{figure1}}
\end{figure}

To investigate the SSE in the all-oxide system, a longitudinal configuration is employed in this work \cite{Uchida2010a,Uchida2012}. In Fig. \ref{figure1}(a), we show a schematic illustration of the experimental configuration and the sample structure of the IrO$_2$/YIG bilayer film for measuring the longitudinal SSE (LSSE). The single-crystalline YIG film was grown on a 0.5-mm-thick (111) Gd$_3$Ga$_5$O$_{12}$ (GGG) substrate by using a liquid phase epitaxy method. The thickness of the YIG film is about 4.5 $\mu$m. A 30-nm-thick IrO$_2$ film was then deposited on the YIG film by using an rf-sputtering method at room temperature. As shown in Fig. \ref{figure1}(b), the light transmittance of the 30-nm-thick IrO$_2$ film is much higher than that of a conventional 10-nm-thick Pt film in the visible light range.

The LSSE measurements were performed by using an experimental setup similar to that described in Ref. \cite{Uchida2012}. The IrO$_2$/YIG sample with the size of $2\times 6$ mm$^2$ was sandwiched between two AlN heat baths of which the temperatures were stabilized to $300~\textrm{K} + \Delta T$ and 300 K, where the temperature of the heat bath connected to the top of the IrO$_2$ layer is higher than that connected to the bottom of the GGG substrate. The temperature difference $\Delta T$ was generated by using a Peltier thermoelectric module and detected by using thermocouples.

The temperature gradient across the IrO$_2$/YIG interface induces a spin current in the IrO$_2$ layer along the direction normal to the interface if the magnetic moments in YIG and conduction-carriers' spins in IrO$_2$ are coupled via the spin-mixing conductance. This spin current is converted into an electric field $\textbf{E}_{\rm{ISHE}}$ by the ISHE in the IrO$_2$ layer along the direction determined by the following relation:
\begin{equation}
\bm{{\rm{E}}}_{\rm{ISHE}} =g_{\rm{r}}^{\rm{\uparrow \downarrow }}\frac {\theta_{\rm{SHE}}\lambda\rho}{d}\bm{{\rm{j}}}_{ \rm{s }}\times\frac { \bm{{\rm{M}}} }{\left|  \bm{{\rm{M}}}\right|}, \label{eq1}
\end{equation}
where $\textbf{j}_{\rm{s}}$, $\theta_{\rm{SHE}}$, $\rho$, $\lambda$, and $d$ denote the spatial direction of the thermally generated spin current, spin-Hall angle, resistivity, spin-diffusion length, and thickness of the IrO$_2$ film, respectively. $g_{\rm{r}}^{\rm{\uparrow \downarrow }}$ is the real part of the spin-mixing conductance at the IrO$_2$/YIG interface and $\textbf{M}$ is the magnetization vector of the YIG film. In Eq. (\ref{eq1}), we neglect the diffusion term of the spin current in the IrO$_2$ film because $d$ ($= 30~\textrm{nm}$) of our sample is much greater than $\lambda$ of IrO$_2$ ($= 3.8~\textrm{nm}$ \cite{Fujiwara2013}). The LSSE-induced $\textbf{E}_{\rm{ISHE}}$ in the IrO$_2$ layer can be detected as an electric voltage signal $V_{\rm{ISHE}} = E_{\rm{ISHE}}l$ with $E_{\rm{ISHE}}$ and $l$ respectively being the magnitude of $\textbf{E}_{\rm{ISHE}}$ and the effective sample length. To measure $V_{\rm{ISHE}}$ induced by the LSSE in the IrO$_2$/YIG sample, two silver-paste electrodes were attached to the ends of the IrO$_2$ layer with the interval of $l=4~\textrm{mm}$ and the electric voltage difference $V$ between the two electrodes was measured with sweeping an external magnetic field $H$ at various values of $\Delta T$.

\begin{figure}
\centering
\begin{minipage}[b]{0.5\textwidth}
\centering
\includegraphics[width=3in]{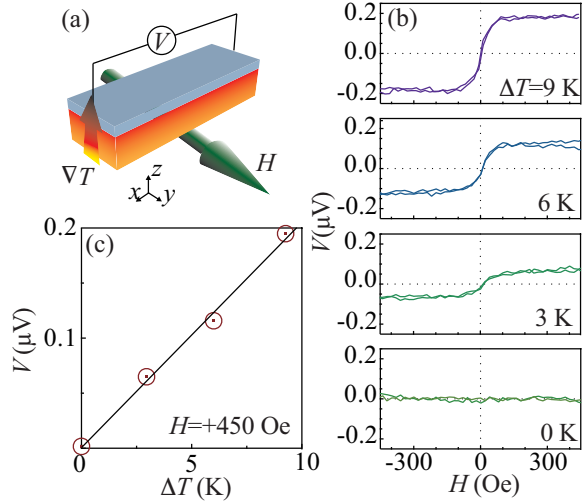}
\end{minipage}
\caption{(a) A schematic illustration of the IrO$_2$/YIG sample. (b) $H$ dependence of the electric voltage $V$ in the IrO$_2$/YIG sample for various values of $\Delta T$. (c) $\Delta T$ dependence of $V$ at $H = +450~\textrm{Oe}$ in the IrO$_2$/YIG sample, measured when $\nabla T$ was applied along the $+z$ direction. \label{figure2}}
\end{figure}

Figure \ref{figure2}(b) shows the $H$ dependence of $V$ for various values of $\Delta T$ for the IrO$_2$/YIG sample, measured when the magnetic field was applied along the $y$ direction. As $\Delta T$ increases, a clear voltage signal was found to appear in the IrO$_2$ layer in response to the magnetization reversal of the YIG layer, while no signal was observed at $\Delta T=0$ K. The voltage signal is proportional to $\Delta T$, and its linear fitting line shows that the thermopower in the IrO$_2$/YIG sample is $V/\Delta T=0.021~\mu$VK$^{-1}$ [see Fig. \ref{figure2}(c)]. These results indicate that the voltage signal is attributed to the ISHE in the IrO$_2$ layer generated by the LSSE in the YIG layer.

\begin{figure}
\centering
\begin{minipage}[b]{0.5\textwidth}
\centering
\includegraphics[width=3in]{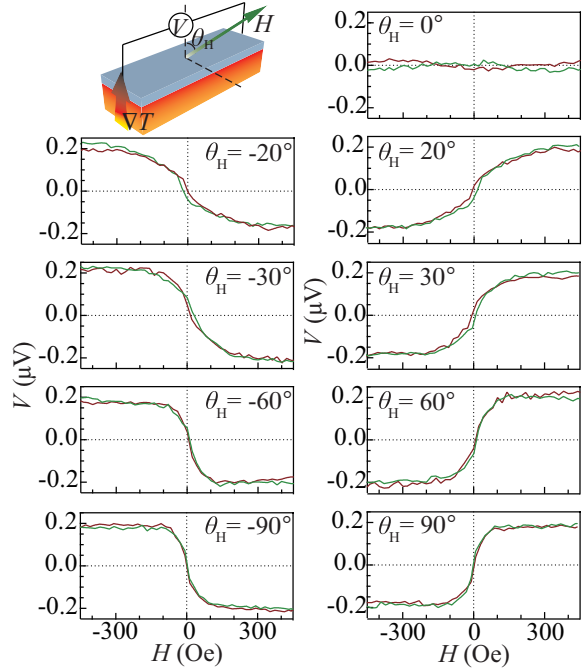}
\end{minipage}
\caption{$H$ dependence of $V$ in the IrO$_2$/YIG sample at $\Delta T=9$ K for various values of the out-of-plane field angle $\theta_{\rm{H}}$. 
\label{figure3}}
\end{figure}

\begin{figure}
\centering
\begin{minipage}[b]{0.5\textwidth}
\centering
\includegraphics[width=3in]{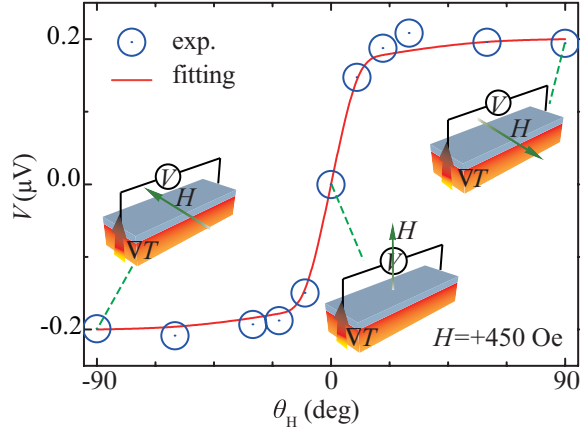}
\end{minipage}
\caption{$\theta_{\rm{H}}$ dependence of $V$ in the IrO$_2$/YIG sample at $H = +450~\textrm{Oe}$. The solid line was obtained by fitting the experimental data with Eq. (\ref{eq1}) combined with static demagnetizing fields in the YIG film. \label{figure4}}
\end{figure}

To further confirm the origin of the thermoelectric voltage in the IrO$_2$/YIG sample, we measured $V$ with changing the angle of the external mangetic field. Figure \ref{figure3} shows the $H$ dependence of $V$ for the IrO$_2$/YIG sample at $\Delta T = 9~\textrm{K}$ for various values of the out-of-plane field angle $\theta_{\rm{H}}$, where $\theta_{\rm{H}}$ is defined as in the upper left panel of Fig. \ref{figure3}. When $\theta_{\rm{H}}\neq 0^{\circ}$, finite $V$ signals appear and their magnitude and sign systematically change with $\theta_{\rm{H}}$. In contrast, no $V$ signal was observed when the magnetic field is perpendicular to the film surface: $\theta_{\rm{H}}=0^{\circ}$. The $\theta_{\rm{H}}$ dependence of $V$ in the IrO$_2$/YIG sample is well reproduced by Eq. (\ref{eq1}) combined with static demagnetizing fields in the YIG film (see Fig. \ref{figure4}), consistent with the symmetry of the ISHE voltage induced by the LSSE. 

\begin{figure}
\centering
\begin{minipage}[b]{0.5\textwidth}
\centering
\includegraphics[width=2in]{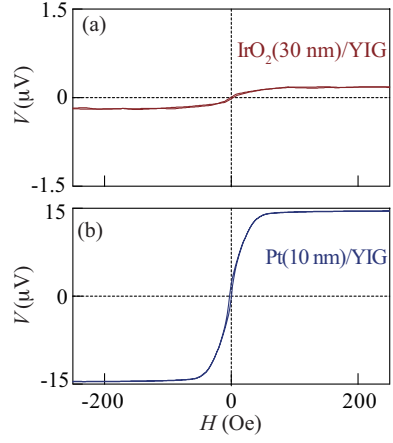}
\end{minipage}
\caption{$H$ dependence of $V$ in the IrO$_2$/YIG (a) and Pt/YIG (b) samples at $\Delta T = 9~\textrm{K}$. The thickness of the IrO$_2$ (Pt) layer of the IrO$_2$/YIG (Pt/YIG) sample is 30 nm (10 nm). \label{figure5}}
\end{figure}

In Fig. \ref{figure5}, we show the comparison of the LSSE signals between the IrO$_2$/YIG sample and conventional Pt/YIG sample. The Pt/YIG sample was prepared by sputtering a 10-nm-thick Pt film on YIG/GGG wafer, where the YIG films for the IrO$_2$/YIG and Pt/YIG samples were grown at the same time. Both the samples have the same size ($2\times 6$ mm$^2$) and the LSSE measurements were carried out in the same condition at $\Delta T=9~\textrm{K}$. The magnitude of the LSSE signal in the IrO$_2$/YIG sample was found to be 68 times smaller than that in the Pt/YIG sample. Such a small LSSE signal in the IrO$_2$/YIG sample is attributed not only to the thicker thickness of the IrO$_2$ layer but also to the small spin-mixing conductance at the IrO$_2$/YIG interface; by using Eq. (\ref{eq1}) with $\theta_{\rm{SHE}} \cdot \lambda = 0.152~\textrm{nm}$ \cite{Fujiwara2013} ($0.188~\textrm{nm}$ \cite{Rojas-Sanchez2014}) and $\rho = 2.7\times10^{-4}\ \Omega$cm ($4.5\times10^{-5}\ \Omega$cm) for the IrO$_2$ (Pt) film and $g_{\rm{r}}^{\rm{\uparrow \downarrow }}=1.3\times10^{18}\ \rm{m}^{-2}$ at the Pt/YIG interface \cite{Qiu2013}, the spin mixing conductance at the IrO$_2$/YIG interface is estimated to be $g_{\rm{r}}^{\rm{\uparrow \downarrow }}=1.2\times10^{16}\ \rm{m}^{-2}$. Therefore, improvement of the spin-mixing conductance at the conductive-oxide/magnetic-insulator interface is indispensable to realizing efficient all-oxide SSE devices, which may be achieved, for example, by improving crystalline structure of the interface by annealing treatment\cite{Qiu2013}, by modulating carrier density in the conductive oxide layer, and by inserting magnetic interlayers between the conductive-oxide and magnetic-insulator layers\cite{Kikuchi2015}.

In summary, we measured the longitudinal spin-Seebeck effect (LSSE) in the all-oxide IrO$_2$/Y$_3$Fe$_5$O$_{12}$ (YIG) bilayer film. The temperature-difference, magnetic-field, and field-angle dependences of the thermoelectric voltage in the IrO$_2$/YIG sample are well consistent with the characteristics of the inverse spin-Hall effect in the IrO$_2$ layer induced by the LSSE in the YIG layer. The LSSE voltage in the IrO$_2$/YIG sample was observed to be much smaller than that in a conventional Pt/YIG sample, which may be attributed to the small spin-mixing conductance at the IrO$_2$/YIG interface if the spin-Hall angle and spin-diffusion length of our IrO$_2$ film are assumed to be comparable to those reported by previous studies. Although an all-oxide system is one of promising candidates for realizing low-cost, stable, and transparent LSSE thermospin devices, major improvement of the spin-mixing conductance at conductive-oxide/magnetic-insulator interfaces is necessary.

The authors thank A. Kirihara and M. Ishida for valuable discussions. This work was supported by PRESTO ``Phase Interfaces for Highly Efficient Energy Utilization'', Strategic International Cooperative Program ASPIMATT from JST, Japan, Grant-in-Aid for Young Scientists (A) (25707029), Grant-in-Aid for Challenging Exploratory Research (26600067), Grant-in-Aid for Scientific Research (A) (24244051), Grant-in-Aid for Scientific Research on Innovative Areas ``Nano Spin Conversion Science'' (26103005) from MEXT, Japan, and NEC Corporation. \par


\begin{thebibliography}{30}
\expandafter\ifx\csname natexlab\endcsname\relax\def\natexlab#1{#1}\fi
\expandafter\ifx\csname bibnamefont\endcsname\relax
  \def\bibnamefont#1{#1}\fi
\expandafter\ifx\csname bibfnamefont\endcsname\relax
  \def\bibfnamefont#1{#1}\fi
\expandafter\ifx\csname citenamefont\endcsname\relax
  \def\citenamefont#1{#1}\fi
\expandafter\ifx\csname url\endcsname\relax
  \def\url#1{\texttt{#1}}\fi
\expandafter\ifx\csname urlprefix\endcsname\relax\def\urlprefix{URL }\fi
\providecommand{\bibinfo}[2]{#2}
\providecommand{\eprint}[2][]{\url{#2}}

\bibitem[{\citenamefont{Weiler et~al.}(2012)\citenamefont{Weiler, Althammer,
  Czeschka, Huebl, Wagner, Opel, Imort, Reiss, Thomas, Gross
  et~al.}}]{Weiler2012}
\bibinfo{author}{\bibfnamefont{M.}~\bibnamefont{Weiler}},
  \bibinfo{author}{\bibfnamefont{M.}~\bibnamefont{Althammer}},
  \bibinfo{author}{\bibfnamefont{F.~D.} \bibnamefont{Czeschka}},
  \bibinfo{author}{\bibfnamefont{H.}~\bibnamefont{Huebl}},
  \bibinfo{author}{\bibfnamefont{M.~S.} \bibnamefont{Wagner}},
  \bibinfo{author}{\bibfnamefont{M.}~\bibnamefont{Opel}},
  \bibinfo{author}{\bibfnamefont{I.~M.} \bibnamefont{Imort}},
  \bibinfo{author}{\bibfnamefont{G.}~\bibnamefont{Reiss}},
  \bibinfo{author}{\bibfnamefont{A.}~\bibnamefont{Thomas}},
  \bibinfo{author}{\bibfnamefont{R.}~\bibnamefont{Gross}},
  \bibnamefont{et~al.}, \bibinfo{journal}{Physical Review Letters}
  \textbf{\bibinfo{volume}{108}}, \bibinfo{pages}{106602}
  (\bibinfo{year}{2012}).

\bibitem[{\citenamefont{Uchida et~al.}(2008)\citenamefont{Uchida, Takahashi,
  Harii, Ieda, Koshibae, Ando, Maekawa, and Saitoh}}]{Uchida2008}
\bibinfo{author}{\bibfnamefont{K.}~\bibnamefont{Uchida}},
  \bibinfo{author}{\bibfnamefont{S.}~\bibnamefont{Takahashi}},
  \bibinfo{author}{\bibfnamefont{K.}~\bibnamefont{Harii}},
  \bibinfo{author}{\bibfnamefont{J.}~\bibnamefont{Ieda}},
  \bibinfo{author}{\bibfnamefont{W.}~\bibnamefont{Koshibae}},
  \bibinfo{author}{\bibfnamefont{K.}~\bibnamefont{Ando}},
  \bibinfo{author}{\bibfnamefont{S.}~\bibnamefont{Maekawa}}, \bibnamefont{and}
  \bibinfo{author}{\bibfnamefont{E.}~\bibnamefont{Saitoh}},
  \bibinfo{journal}{Nature} \textbf{\bibinfo{volume}{455}},
  \bibinfo{pages}{778} (\bibinfo{year}{2008}).

\bibitem[{\citenamefont{Jaworski et~al.}(2010)\citenamefont{Jaworski, Yang,
  Mack, Awschalom, Heremans, and Myers}}]{Jaworski2010}
\bibinfo{author}{\bibfnamefont{C.~M.} \bibnamefont{Jaworski}},
  \bibinfo{author}{\bibfnamefont{J.}~\bibnamefont{Yang}},
  \bibinfo{author}{\bibfnamefont{S.}~\bibnamefont{Mack}},
  \bibinfo{author}{\bibfnamefont{D.~D.} \bibnamefont{Awschalom}},
  \bibinfo{author}{\bibfnamefont{J.~P.} \bibnamefont{Heremans}},
  \bibnamefont{and} \bibinfo{author}{\bibfnamefont{R.~C.} \bibnamefont{Myers}},
  \bibinfo{journal}{Nature Materials} \textbf{\bibinfo{volume}{9}},
  \bibinfo{pages}{898} (\bibinfo{year}{2010}).

\bibitem[{\citenamefont{Qu et~al.}(2013)\citenamefont{Qu, Huang, Hu, Wu, and
  Chien}}]{Qu2013}
\bibinfo{author}{\bibfnamefont{D.}~\bibnamefont{Qu}},
  \bibinfo{author}{\bibfnamefont{S.~Y.} \bibnamefont{Huang}},
  \bibinfo{author}{\bibfnamefont{J.}~\bibnamefont{Hu}},
  \bibinfo{author}{\bibfnamefont{R.}~\bibnamefont{Wu}}, \bibnamefont{and}
  \bibinfo{author}{\bibfnamefont{C.~L.} \bibnamefont{Chien}},
  \bibinfo{journal}{Physical Review Letters} \textbf{\bibinfo{volume}{110}},
  \bibinfo{pages}{067206} (\bibinfo{year}{2013}).

\bibitem[{\citenamefont{Agrawal et~al.}(2014)\citenamefont{Agrawal, Vasyuchka,
  Serga, Kirihara, Pirro, Langner, Jungfleisch, Chumak, Papaioannou, and
  Hillebrands}}]{Agrawal2014}
\bibinfo{author}{\bibfnamefont{M.}~\bibnamefont{Agrawal}},
  \bibinfo{author}{\bibfnamefont{V.~I.} \bibnamefont{Vasyuchka}},
  \bibinfo{author}{\bibfnamefont{A.~A.} \bibnamefont{Serga}},
  \bibinfo{author}{\bibfnamefont{A.}~\bibnamefont{Kirihara}},
  \bibinfo{author}{\bibfnamefont{P.}~\bibnamefont{Pirro}},
  \bibinfo{author}{\bibfnamefont{T.}~\bibnamefont{Langner}},
  \bibinfo{author}{\bibfnamefont{M.~B.} \bibnamefont{Jungfleisch}},
  \bibinfo{author}{\bibfnamefont{a.~V.} \bibnamefont{Chumak}},
  \bibinfo{author}{\bibfnamefont{E.~T.} \bibnamefont{Papaioannou}},
  \bibnamefont{and}
  \bibinfo{author}{\bibfnamefont{B.}~\bibnamefont{Hillebrands}},
  \bibinfo{journal}{Physical Review B} \textbf{\bibinfo{volume}{89}},
  \bibinfo{pages}{224414} (\bibinfo{year}{2014}).

\bibitem[{\citenamefont{Rezende et~al.}(2014)\citenamefont{Rezende,
  Rodr\'{\i}guez-Su\'{a}rez, Cunha, Rodrigues, Machado, {Fonseca Guerra},
  {Lopez Ortiz}, and Azevedo}}]{Rezende2014}
\bibinfo{author}{\bibfnamefont{S.~M.} \bibnamefont{Rezende}},
  \bibinfo{author}{\bibfnamefont{R.~L.}
  \bibnamefont{Rodr\'{\i}guez-Su\'{a}rez}},
  \bibinfo{author}{\bibfnamefont{R.~O.} \bibnamefont{Cunha}},
  \bibinfo{author}{\bibfnamefont{A.~R.} \bibnamefont{Rodrigues}},
  \bibinfo{author}{\bibfnamefont{F.~L.~A.} \bibnamefont{Machado}},
  \bibinfo{author}{\bibfnamefont{G.~A.} \bibnamefont{{Fonseca Guerra}}},
  \bibinfo{author}{\bibfnamefont{J.~C.} \bibnamefont{{Lopez Ortiz}}},
  \bibnamefont{and} \bibinfo{author}{\bibfnamefont{A.}~\bibnamefont{Azevedo}},
  \bibinfo{journal}{Physical Review B} \textbf{\bibinfo{volume}{89}},
  \bibinfo{pages}{014416} (\bibinfo{year}{2014}).

\bibitem[{\citenamefont{{R. Ramos} et~al.}(2013)\citenamefont{{R. Ramos},
  Kikkawa, Uchida, Adachi, Lucas, Aguirre, Algarabel, On, Maekawa, Saitoh
  et~al.}}]{R.Ramos2013}
\bibinfo{author}{\bibnamefont{{R. Ramos}}},
  \bibinfo{author}{\bibfnamefont{T.}~\bibnamefont{Kikkawa}},
  \bibinfo{author}{\bibfnamefont{K.}~\bibnamefont{Uchida}},
  \bibinfo{author}{\bibfnamefont{H.}~\bibnamefont{Adachi}},
  \bibinfo{author}{\bibfnamefont{I.}~\bibnamefont{Lucas}},
  \bibinfo{author}{\bibfnamefont{M.~H.} \bibnamefont{Aguirre}},
  \bibinfo{author}{\bibfnamefont{P.}~\bibnamefont{Algarabel}},
  \bibinfo{author}{\bibfnamefont{L.~M.} \bibnamefont{On}},
  \bibinfo{author}{\bibfnamefont{S.}~\bibnamefont{Maekawa}},
  \bibinfo{author}{\bibfnamefont{E.}~\bibnamefont{Saitoh}},
  \bibnamefont{and} \bibinfo{author}{\bibfnamefont{M. R.}~\bibnamefont{Ibarra}},
  \bibinfo{journal}{Appl. Phys. Lett.}
  \textbf{\bibinfo{volume}{102}}, \bibinfo{pages}{072413}
  (\bibinfo{year}{2013}).

\bibitem[{\citenamefont{Meier et~al.}(2013)\citenamefont{Meier, Kuschel, Shen,
  Gupta, Kikkawa, Uchida, Saitoh, Schmalhorst, and Reiss}}]{Meier2013}
\bibinfo{author}{\bibfnamefont{D.}~\bibnamefont{Meier}},
  \bibinfo{author}{\bibfnamefont{T.}~\bibnamefont{Kuschel}},
  \bibinfo{author}{\bibfnamefont{L.}~\bibnamefont{Shen}},
  \bibinfo{author}{\bibfnamefont{A.}~\bibnamefont{Gupta}},
  \bibinfo{author}{\bibfnamefont{T.}~\bibnamefont{Kikkawa}},
  \bibinfo{author}{\bibfnamefont{K.}~\bibnamefont{Uchida}},
  \bibinfo{author}{\bibfnamefont{E.}~\bibnamefont{Saitoh}},
  \bibinfo{author}{\bibfnamefont{J.~M.} \bibnamefont{Schmalhorst}},
  \bibnamefont{and} \bibinfo{author}{\bibfnamefont{G.}~\bibnamefont{Reiss}},
  \bibinfo{journal}{Physical Review B} \textbf{\bibinfo{volume}{87}},
  \bibinfo{pages}{054421} (\bibinfo{year}{2013}).

\bibitem[{\citenamefont{Uchida et~al.}(2010{\natexlab{a}})\citenamefont{Uchida,
  Xiao, Adachi, Ohe, Takahashi, Ieda, Ota, Kajiwara, Umezawa, Kawai
  et~al.}}]{Uchida2010}
\bibinfo{author}{\bibfnamefont{K.}~\bibnamefont{Uchida}},
  \bibinfo{author}{\bibfnamefont{J.}~\bibnamefont{Xiao}},
  \bibinfo{author}{\bibfnamefont{H.}~\bibnamefont{Adachi}},
  \bibinfo{author}{\bibfnamefont{J.}~\bibnamefont{Ohe}},
  \bibinfo{author}{\bibfnamefont{S.}~\bibnamefont{Takahashi}},
  \bibinfo{author}{\bibfnamefont{J.}~\bibnamefont{Ieda}},
  \bibinfo{author}{\bibfnamefont{T.}~\bibnamefont{Ota}},
  \bibinfo{author}{\bibfnamefont{Y.}~\bibnamefont{Kajiwara}},
  \bibinfo{author}{\bibfnamefont{H.}~\bibnamefont{Umezawa}},
  \bibinfo{author}{\bibfnamefont{H.}~\bibnamefont{Kawai}},
  \bibinfo{author}{\bibfnamefont{G. E. W.}~\bibnamefont{Bauer}},
  \bibinfo{author}{\bibfnamefont{S.}~\bibnamefont{Maekawa}},
  \bibnamefont{and} \bibinfo{author}{\bibfnamefont{E.}~\bibnamefont{Saitoh}},
  \bibinfo{journal}{Nature Materials}
  \textbf{\bibinfo{volume}{9}}, \bibinfo{pages}{894}
  (\bibinfo{year}{2010}{\natexlab{a}}).

\bibitem[{\citenamefont{Uchida et~al.}(2010{\natexlab{b}})\citenamefont{Uchida,
  Adachi, Ota, Nakayama, Maekawa, and Saitoh}}]{Uchida2010a}
\bibinfo{author}{\bibfnamefont{K.}~\bibnamefont{Uchida}},
  \bibinfo{author}{\bibfnamefont{H.}~\bibnamefont{Adachi}},
  \bibinfo{author}{\bibfnamefont{T.}~\bibnamefont{Ota}},
  \bibinfo{author}{\bibfnamefont{H.}~\bibnamefont{Nakayama}},
  \bibinfo{author}{\bibfnamefont{S.}~\bibnamefont{Maekawa}}, \bibnamefont{and}
  \bibinfo{author}{\bibfnamefont{E.}~\bibnamefont{Saitoh}},
  \bibinfo{journal}{Applied Physics Letters} \textbf{\bibinfo{volume}{97}},
  \bibinfo{pages}{172505} (\bibinfo{year}{2010}{\natexlab{b}}).

\bibitem[{\citenamefont{Uchida et~al.}(2012)\citenamefont{Uchida, Ota, Adachi,
  Xiao, Nonaka, Kajiwara, Bauer, Maekawa, and Saitoh}}]{Uchida2012}
\bibinfo{author}{\bibfnamefont{K.}~\bibnamefont{Uchida}},
  \bibinfo{author}{\bibfnamefont{T.}~\bibnamefont{Ota}},
  \bibinfo{author}{\bibfnamefont{H.}~\bibnamefont{Adachi}},
  \bibinfo{author}{\bibfnamefont{J.}~\bibnamefont{Xiao}},
  \bibinfo{author}{\bibfnamefont{T.}~\bibnamefont{Nonaka}},
  \bibinfo{author}{\bibfnamefont{Y.}~\bibnamefont{Kajiwara}},
  \bibinfo{author}{\bibfnamefont{G.~E.~W.} \bibnamefont{Bauer}},
  \bibinfo{author}{\bibfnamefont{S.}~\bibnamefont{Maekawa}}, \bibnamefont{and}
  \bibinfo{author}{\bibfnamefont{E.}~\bibnamefont{Saitoh}},
  \bibinfo{journal}{Journal of Applied Physics} \textbf{\bibinfo{volume}{111}},
  \bibinfo{pages}{103903} (\bibinfo{year}{2012}).

\bibitem[{\citenamefont{Kikkawa
  et~al.}(2013{\natexlab{a}})\citenamefont{Kikkawa, Uchida, Daimon, Shiomi,
  Adachi, Qiu, Hou, Jin, Maekawa, and Saitoh}}]{Kikkawa2013}
\bibinfo{author}{\bibfnamefont{T.}~\bibnamefont{Kikkawa}},
  \bibinfo{author}{\bibfnamefont{K.}~\bibnamefont{Uchida}},
  \bibinfo{author}{\bibfnamefont{S.}~\bibnamefont{Daimon}},
  \bibinfo{author}{\bibfnamefont{Y.}~\bibnamefont{Shiomi}},
  \bibinfo{author}{\bibfnamefont{H.}~\bibnamefont{Adachi}},
  \bibinfo{author}{\bibfnamefont{Z.}~\bibnamefont{Qiu}},
  \bibinfo{author}{\bibfnamefont{D.}~\bibnamefont{Hou}},
  \bibinfo{author}{\bibfnamefont{X.~F.} \bibnamefont{Jin}},
  \bibinfo{author}{\bibfnamefont{S.}~\bibnamefont{Maekawa}}, \bibnamefont{and}
  \bibinfo{author}{\bibfnamefont{E.}~\bibnamefont{Saitoh}},
  \bibinfo{journal}{Physical Review B} \textbf{\bibinfo{volume}{88}},
  \bibinfo{pages}{214403} (\bibinfo{year}{2013}{\natexlab{a}}).

\bibitem[{\citenamefont{Kikkawa
  et~al.}(2013{\natexlab{b}})\citenamefont{Kikkawa, Uchida, Shiomi, Qiu, Hou,
  Tian, Nakayama, Jin, and Saitoh}}]{Kikkawa2013a}
\bibinfo{author}{\bibfnamefont{T.}~\bibnamefont{Kikkawa}},
  \bibinfo{author}{\bibfnamefont{K.}~\bibnamefont{Uchida}},
  \bibinfo{author}{\bibfnamefont{Y.}~\bibnamefont{Shiomi}},
  \bibinfo{author}{\bibfnamefont{Z.}~\bibnamefont{Qiu}},
  \bibinfo{author}{\bibfnamefont{D.}~\bibnamefont{Hou}},
  \bibinfo{author}{\bibfnamefont{D.}~\bibnamefont{Tian}},
  \bibinfo{author}{\bibfnamefont{H.}~\bibnamefont{Nakayama}},
  \bibinfo{author}{\bibfnamefont{X.~F.} \bibnamefont{Jin}}, \bibnamefont{and}
  \bibinfo{author}{\bibfnamefont{E.}~\bibnamefont{Saitoh}},
  \bibinfo{journal}{Physical Review Letters} \textbf{\bibinfo{volume}{110}},
  \bibinfo{pages}{067207} (\bibinfo{year}{2013}{\natexlab{b}}).

\bibitem[{\citenamefont{Kirihara et~al.}(2012)\citenamefont{Kirihara, Uchida,
  Kajiwara, Ishida, Nakamura, Manako, Saitoh, and Yorozu}}]{Kirihara2012}
\bibinfo{author}{\bibfnamefont{A.}~\bibnamefont{Kirihara}},
  \bibinfo{author}{\bibfnamefont{K.}~\bibnamefont{Uchida}},
  \bibinfo{author}{\bibfnamefont{Y.}~\bibnamefont{Kajiwara}},
  \bibinfo{author}{\bibfnamefont{M.}~\bibnamefont{Ishida}},
  \bibinfo{author}{\bibfnamefont{Y.}~\bibnamefont{Nakamura}},
  \bibinfo{author}{\bibfnamefont{T.}~\bibnamefont{Manako}},
  \bibinfo{author}{\bibfnamefont{E.}~\bibnamefont{Saitoh}}, \bibnamefont{and}
  \bibinfo{author}{\bibfnamefont{S.}~\bibnamefont{Yorozu}},
  \bibinfo{journal}{Nature Materials} \textbf{\bibinfo{volume}{11}},
  \bibinfo{pages}{686} (\bibinfo{year}{2012}).

\bibitem[{\citenamefont{Roschewsky et~al.}(2014)\citenamefont{Roschewsky,
  Schreier, Kamra, Schade, Ganzhorn, Meyer, Huebl, Gross, and
  Goennenwein}}]{Roschewsky2014}
\bibinfo{author}{\bibfnamefont{N.}~\bibnamefont{Roschewsky}},
  \bibinfo{author}{\bibfnamefont{M.}~\bibnamefont{Schreier}},
  \bibinfo{author}{\bibfnamefont{A.}~\bibnamefont{Kamra}},
  \bibinfo{author}{\bibfnamefont{F.}~\bibnamefont{Schade}},
  \bibinfo{author}{\bibfnamefont{K.}~\bibnamefont{Ganzhorn}},
  \bibinfo{author}{\bibfnamefont{S.}~\bibnamefont{Meyer}},
  \bibinfo{author}{\bibfnamefont{H.}~\bibnamefont{Huebl}},
  \bibinfo{author}{\bibfnamefont{R.}~\bibnamefont{Gross}}, \bibnamefont{and}
  \bibinfo{author}{\bibfnamefont{S.~T.~B.} \bibnamefont{Goennenwein}},
  \bibinfo{journal}{Applied Physics Letters} \textbf{\bibinfo{volume}{104}},
  \bibinfo{pages}{202410} (\bibinfo{year}{2014}).

\bibitem[{\citenamefont{Uchida et~al.}(2014{\natexlab{a}})\citenamefont{Uchida,
  Ishida, Kikkawa, Kirihara, Murakami, and Saitoh}}]{Uchida2014a}
\bibinfo{author}{\bibfnamefont{K.}~\bibnamefont{Uchida}},
  \bibinfo{author}{\bibfnamefont{M.}~\bibnamefont{Ishida}},
  \bibinfo{author}{\bibfnamefont{T.}~\bibnamefont{Kikkawa}},
  \bibinfo{author}{\bibfnamefont{A.}~\bibnamefont{Kirihara}},
  \bibinfo{author}{\bibfnamefont{T.}~\bibnamefont{Murakami}}, \bibnamefont{and}
  \bibinfo{author}{\bibfnamefont{E.}~\bibnamefont{Saitoh}},
  \bibinfo{journal}{Journal of Physics: Condensed matter}
  \textbf{\bibinfo{volume}{26}}, \bibinfo{pages}{343202}
  (\bibinfo{year}{2014}{\natexlab{a}}).

\bibitem[{\citenamefont{Uchida et~al.}(2014{\natexlab{b}})\citenamefont{Uchida,
  Kikkawa, Miura, Shiomi, and Saitoh}}]{PhysRevX.4.041023}
\bibinfo{author}{\bibfnamefont{K.}~\bibnamefont{Uchida}},
  \bibinfo{author}{\bibfnamefont{T.}~\bibnamefont{Kikkawa}},
  \bibinfo{author}{\bibfnamefont{A.}~\bibnamefont{Miura}},
  \bibinfo{author}{\bibfnamefont{J.}~\bibnamefont{Shiomi}}, \bibnamefont{and}
  \bibinfo{author}{\bibfnamefont{E.}~\bibnamefont{Saitoh}},
  \bibinfo{journal}{Phys. Rev. X} \textbf{\bibinfo{volume}{4}},
  \bibinfo{pages}{041023} (\bibinfo{year}{2014}{\natexlab{b}}).

\bibitem[{\citenamefont{Schreier et~al.}(2013)\citenamefont{Schreier,
  Roschewsky, Dobler, Meyer, Huebl, Gross, and Goennenwein}}]{Schreier2013a}
\bibinfo{author}{\bibfnamefont{M.}~\bibnamefont{Schreier}},
  \bibinfo{author}{\bibfnamefont{N.}~\bibnamefont{Roschewsky}},
  \bibinfo{author}{\bibfnamefont{E.}~\bibnamefont{Dobler}},
  \bibinfo{author}{\bibfnamefont{S.}~\bibnamefont{Meyer}},
  \bibinfo{author}{\bibfnamefont{H.}~\bibnamefont{Huebl}},
  \bibinfo{author}{\bibfnamefont{R.}~\bibnamefont{Gross}}, \bibnamefont{and}
  \bibinfo{author}{\bibfnamefont{S.~T.~B.} \bibnamefont{Goennenwein}},
  \bibinfo{journal}{Applied Physics Letters} \textbf{\bibinfo{volume}{103}},
  \bibinfo{pages}{242404} (\bibinfo{year}{2013}).

\bibitem[{\citenamefont{Uchida et~al.}(2013)\citenamefont{Uchida, Nonaka,
  Kikkawa, Kajiwara, and Saitoh}}]{Uchida2013a}
\bibinfo{author}{\bibfnamefont{K.}~\bibnamefont{Uchida}},
  \bibinfo{author}{\bibfnamefont{T.}~\bibnamefont{Nonaka}},
  \bibinfo{author}{\bibfnamefont{T.}~\bibnamefont{Kikkawa}},
  \bibinfo{author}{\bibfnamefont{Y.}~\bibnamefont{Kajiwara}}, \bibnamefont{and}
  \bibinfo{author}{\bibfnamefont{E.}~\bibnamefont{Saitoh}},
  \bibinfo{journal}{Physical Review B} \textbf{\bibinfo{volume}{87}},
  \bibinfo{pages}{104412} (\bibinfo{year}{2013}).

\bibitem[{\citenamefont{Bauer et~al.}(2012)\citenamefont{Bauer, Saitoh, and van
  Wees}}]{Bauer2012}
\bibinfo{author}{\bibfnamefont{G.~E.~W.} \bibnamefont{Bauer}},
  \bibinfo{author}{\bibfnamefont{E.}~\bibnamefont{Saitoh}}, \bibnamefont{and}
  \bibinfo{author}{\bibfnamefont{B.~J.} \bibnamefont{van Wees}},
  \bibinfo{journal}{Nature Materials} \textbf{\bibinfo{volume}{11}},
  \bibinfo{pages}{391} (\bibinfo{year}{2012}).

\bibitem[{\citenamefont{Geller and Gilleo}(1957)}]{Geller1957}
\bibinfo{author}{\bibfnamefont{S.}~\bibnamefont{Geller}} \bibnamefont{and}
  \bibinfo{author}{\bibfnamefont{M.~A.} \bibnamefont{Gilleo}},
  \bibinfo{journal}{Acta Crystallographica} \textbf{\bibinfo{volume}{10}},
  \bibinfo{pages}{239} (\bibinfo{year}{1957}).

\bibitem[{\citenamefont{Cherepanov et~al.}(1993)\citenamefont{Cherepanov,
  Kolokolov, and L'vov}}]{Cherepanov1993}
\bibinfo{author}{\bibfnamefont{V.}~\bibnamefont{Cherepanov}},
  \bibinfo{author}{\bibfnamefont{I.}~\bibnamefont{Kolokolov}},
  \bibnamefont{and} \bibinfo{author}{\bibfnamefont{V.}~\bibnamefont{L'vov}},
  \bibinfo{journal}{Physics Reports} \textbf{\bibinfo{volume}{229}},
  \bibinfo{pages}{81} (\bibinfo{year}{1993}).

\bibitem[{\citenamefont{Serga et~al.}(2010)\citenamefont{Serga, Chumak, and
  Hillebrands}}]{Serga2010}
\bibinfo{author}{\bibfnamefont{A.~A.} \bibnamefont{Serga}},
  \bibinfo{author}{\bibfnamefont{A.~V.} \bibnamefont{Chumak}},
  \bibnamefont{and}
  \bibinfo{author}{\bibfnamefont{B.}~\bibnamefont{Hillebrands}},
  \bibinfo{journal}{Journal of Physics D-Applied Physics}
  \textbf{\bibinfo{volume}{43}}, \bibinfo{pages}{264002}
  (\bibinfo{year}{2010}).

\bibitem[{\citenamefont{Qiu et~al.}(2013{\natexlab{a}})\citenamefont{Qiu, An,
  Uchida, Hou, Shiomi, Fujikawa, and Saitoh}}]{Qiu2013a}
\bibinfo{author}{\bibfnamefont{Z.}~\bibnamefont{Qiu}},
  \bibinfo{author}{\bibfnamefont{T.}~\bibnamefont{An}},
  \bibinfo{author}{\bibfnamefont{K.}~\bibnamefont{Uchida}},
  \bibinfo{author}{\bibfnamefont{D.}~\bibnamefont{Hou}},
  \bibinfo{author}{\bibfnamefont{Y.}~\bibnamefont{Shiomi}},
  \bibinfo{author}{\bibfnamefont{Y.}~\bibnamefont{Fujikawa}}, \bibnamefont{and}
  \bibinfo{author}{\bibfnamefont{E.}~\bibnamefont{Saitoh}},
  \bibinfo{journal}{Applied Physics Letters} \textbf{\bibinfo{volume}{103}},
  \bibinfo{pages}{182404} (\bibinfo{year}{2013}{\natexlab{a}}).

\bibitem[{\citenamefont{Qiu et~al.}(2012)\citenamefont{Qiu, Kajiwara, Ando,
  Fujikawa, Uchida, Tashiro, Harii, Yoshino, and Saitoh}}]{Qiu2012}
\bibinfo{author}{\bibfnamefont{Z.}~\bibnamefont{Qiu}},
  \bibinfo{author}{\bibfnamefont{Y.}~\bibnamefont{Kajiwara}},
  \bibinfo{author}{\bibfnamefont{K.}~\bibnamefont{Ando}},
  \bibinfo{author}{\bibfnamefont{Y.}~\bibnamefont{Fujikawa}},
  \bibinfo{author}{\bibfnamefont{K.}~\bibnamefont{Uchida}},
  \bibinfo{author}{\bibfnamefont{T.}~\bibnamefont{Tashiro}},
  \bibinfo{author}{\bibfnamefont{K.}~\bibnamefont{Harii}},
  \bibinfo{author}{\bibfnamefont{T.}~\bibnamefont{Yoshino}}, \bibnamefont{and}
  \bibinfo{author}{\bibfnamefont{E.}~\bibnamefont{Saitoh}},
  \bibinfo{journal}{Applied Physics Letters} \textbf{\bibinfo{volume}{100}},
  \bibinfo{pages}{022402} (\bibinfo{year}{2012}).

\bibitem[{\citenamefont{Fujiwara et~al.}(2013)\citenamefont{Fujiwara, Fukuma,
  Matsuno, Idzuchi, Niimi, Otani, and Takagi}}]{Fujiwara2013}
\bibinfo{author}{\bibfnamefont{K.}~\bibnamefont{Fujiwara}},
  \bibinfo{author}{\bibfnamefont{Y.}~\bibnamefont{Fukuma}},
  \bibinfo{author}{\bibfnamefont{J.}~\bibnamefont{Matsuno}},
  \bibinfo{author}{\bibfnamefont{H.}~\bibnamefont{Idzuchi}},
  \bibinfo{author}{\bibfnamefont{Y.}~\bibnamefont{Niimi}},
  \bibinfo{author}{\bibfnamefont{Y.}~\bibnamefont{Otani}}, \bibnamefont{and}
  \bibinfo{author}{\bibfnamefont{H.}~\bibnamefont{Takagi}},
  \bibinfo{journal}{Nature Communications} \textbf{\bibinfo{volume}{4}},
  \bibinfo{pages}{2893} (\bibinfo{year}{2013}).

\bibitem[{\citenamefont{Chalamala et~al.}(1999)\citenamefont{Chalamala, Wei,
  Reuss, Aggarwal, Gnade, Ramesh, Bernhard, Sosa, and Golden}}]{Chalamala1999}
\bibinfo{author}{\bibfnamefont{B.~R.} \bibnamefont{Chalamala}},
  \bibinfo{author}{\bibfnamefont{Y.}~\bibnamefont{Wei}},
  \bibinfo{author}{\bibfnamefont{R.~H.} \bibnamefont{Reuss}},
  \bibinfo{author}{\bibfnamefont{S.}~\bibnamefont{Aggarwal}},
  \bibinfo{author}{\bibfnamefont{B.~E.} \bibnamefont{Gnade}},
  \bibinfo{author}{\bibfnamefont{R.}~\bibnamefont{Ramesh}},
  \bibinfo{author}{\bibfnamefont{J.~M.} \bibnamefont{Bernhard}},
  \bibinfo{author}{\bibfnamefont{E.~D.} \bibnamefont{Sosa}}, \bibnamefont{and}
  \bibinfo{author}{\bibfnamefont{D.~E.} \bibnamefont{Golden}},
  \bibinfo{journal}{Applied Physics Letters} \textbf{\bibinfo{volume}{74}},
  \bibinfo{pages}{1394} (\bibinfo{year}{1999}).

\bibitem[{\citenamefont{Rojas-S\'{a}nchez
  et~al.}(2014)\citenamefont{Rojas-S\'{a}nchez, Reyren, Laczkowski, Savero,
  Attan\'{e}, Deranlot, Jamet, George, Vila, and
  Jaffr\`{e}s}}]{Rojas-Sanchez2014}
\bibinfo{author}{\bibfnamefont{J.~C.} \bibnamefont{Rojas-S\'{a}nchez}},
  \bibinfo{author}{\bibfnamefont{N.}~\bibnamefont{Reyren}},
  \bibinfo{author}{\bibfnamefont{P.}~\bibnamefont{Laczkowski}},
  \bibinfo{author}{\bibfnamefont{W.}~\bibnamefont{Savero}},
  \bibinfo{author}{\bibfnamefont{J.~P.} \bibnamefont{Attan\'{e}}},
  \bibinfo{author}{\bibfnamefont{C.}~\bibnamefont{Deranlot}},
  \bibinfo{author}{\bibfnamefont{M.}~\bibnamefont{Jamet}},
  \bibinfo{author}{\bibfnamefont{J.~M.} \bibnamefont{George}},
  \bibinfo{author}{\bibfnamefont{L.}~\bibnamefont{Vila}}, \bibnamefont{and}
  \bibinfo{author}{\bibfnamefont{H.}~\bibnamefont{Jaffr\`{e}s}},
  \bibinfo{journal}{Physical Review Letters} \textbf{\bibinfo{volume}{112}},
  \bibinfo{pages}{106602} (\bibinfo{year}{2014}).

\bibitem[{\citenamefont{Qiu et~al.}(2013{\natexlab{b}})\citenamefont{Qiu, Ando,
  Uchida, Kajiwara, Takahashi, Nakayama, An, Fujikawa, and Saitoh}}]{Qiu2013}
\bibinfo{author}{\bibfnamefont{Z.}~\bibnamefont{Qiu}},
  \bibinfo{author}{\bibfnamefont{K.}~\bibnamefont{Ando}},
  \bibinfo{author}{\bibfnamefont{K.}~\bibnamefont{Uchida}},
  \bibinfo{author}{\bibfnamefont{Y.}~\bibnamefont{Kajiwara}},
  \bibinfo{author}{\bibfnamefont{R.}~\bibnamefont{Takahashi}},
  \bibinfo{author}{\bibfnamefont{H.}~\bibnamefont{Nakayama}},
  \bibinfo{author}{\bibfnamefont{T.}~\bibnamefont{An}},
  \bibinfo{author}{\bibfnamefont{Y.}~\bibnamefont{Fujikawa}}, \bibnamefont{and}
  \bibinfo{author}{\bibfnamefont{E.}~\bibnamefont{Saitoh}},
  \bibinfo{journal}{Applied Physics Letters} \textbf{\bibinfo{volume}{103}},
  \bibinfo{pages}{092404} (\bibinfo{year}{2013}{\natexlab{b}}).

\bibitem[{\citenamefont{Kikuchi et~al.}(2015)\citenamefont{Kikuchi, Ishida,
  Uchida, Qiu, Murakami, and Saitoh}}]{Kikuchi2015}
\bibinfo{author}{\bibfnamefont{D.}~\bibnamefont{Kikuchi}},
  \bibinfo{author}{\bibfnamefont{M.}~\bibnamefont{Ishida}},
  \bibinfo{author}{\bibfnamefont{K.}~\bibnamefont{Uchida}},
  \bibinfo{author}{\bibfnamefont{Z.}~\bibnamefont{Qiu}},
  \bibinfo{author}{\bibfnamefont{T.}~\bibnamefont{Murakami}}, \bibnamefont{and}
  \bibinfo{author}{\bibfnamefont{E.}~\bibnamefont{Saitoh}},
  \bibinfo{journal}{Appl. Phys. Lett.} \textbf{\bibinfo{volume}{106}},
  \bibinfo{pages}{082401} (\bibinfo{year}{2015}).

\end{thebibliography}



\newpage


\end{document}